\documentclass[twocolumn, iop, apj, twocolappendix, numberedappendix, appendixfloats]{emulateapj}
\usepackage{CJKutf8}
\usepackage{threeparttable}
\usepackage{multirow}
\usepackage{times}
\usepackage{enumitem}
\usepackage{url} 
\usepackage{color}
\usepackage{amsmath,bm}
\usepackage{subfigure}
\usepackage{natbib}
\usepackage{graphicx}
\usepackage{graphics}
\usepackage{hyperref}
\usepackage{amsmath}
\usepackage{bm}
\usepackage{subfigure}
\usepackage{natbib}
\usepackage{array}
\usepackage{booktabs}
\usepackage{lineno}

\shorttitle{The resonances of the Galactic Bar}
\shortauthors{Sun et al.}
\submitted{Submitted to ApJL;\,\,\,\,Accepted March 19, 2024}

\begin{document}

\title{On the chemical and kinematic signatures of the resonances of the Galactic bar as revealed by the LAMOST-APOGEE red clump stars}

\author{Weixiang Sun\textsuperscript{1,3}}
\author{Han Shen\textsuperscript{2}}
\author{Biwei Jiang\textsuperscript{1,3}}
\author{Xiaowei Liu\textsuperscript{2,3}}

\altaffiltext{1}{Department of Astronomy, Beijing Normal University, Beijing 100875, People's Republic of China; {\it sunweixiang@bnu.edu.cn (WXS)}; {\it bjiang@bnu.edu.cn (BWJ)}.}


\altaffiltext{2}{South-Western Institute for Astronomy Research, Yunnan University, Kunming 650500, People's Republic of China; {\it x.liu@ynu.edu.cn (XWL)}.}




\altaffiltext{3}{Corresponding authors}

\begin{abstract}
The Milky Way is widely considered to exhibit features of a rotational bar or quadrupole bar.
In either case, the feature of the resonance of the Galactic bar should be present in the properties of the chemistry and kinematics, over a large area of the disk.
With a sample of over 170,000 red clump (RC) stars from LAMOST-APOGEE data, we attempt to detect the chemical and kinematic signatures of the resonances of the Galactic bar, within 4.0 $\leq$ $R$ $\leq$ 15.0\,kpc and $|Z|$ $\leq$ 3.0\,kpc.
The measurement of the $\Delta$[Fe/H]/$\Delta|Z|$--$R$ with subtracted the global profiles trends, shows that the thin and thick disks values are Cor\_$\Delta$[Fe/H]/$\Delta|Z|$ = 0.010\,$\mathrm{sin}$\,(1.598\,$R$\,+\,2.551) and Cor\_$\Delta$[Fe/H]/$\Delta|Z|$ = 0.006\,$\mathrm{sin}$\,(1.258\,$R$\,$-$\,0.019), respectively.
The analysis of the tilt angle of the velocity ellipsoid indicates that the thin and thick disks are accurately described as $\alpha$ = $\alpha_{0}$ arctan (Z/R), with $\alpha_{0}$ = 0.198\,$\mathrm{sin}$\,(0.853\,$R$\,+\,1.982)\,+\,0.630 and $\alpha_{0}$ = 0.220\,$\mathrm{sin}$\,(0.884\,$R$\,+\,2.012)\,+\,0.679 for thin and thick disks, respectively.
These periodic oscillations in Cor\_$\Delta$[Fe/H]/$\Delta|Z|$ and $\alpha_{0}$ with $R$ appear in both thin and thick disks, are the most likely chemical and kinematic signatures of the resonance of the Galactic bar.
The difference in the phase of the functions of the fitted periodic oscillations for the thin and thick disks may be related to the presence of a second Galactic bar.

\end{abstract}


\section{Introduction}

The Milky Way is a barred galaxy has been inferred first by the gas kinematics \citep[e.g.,][]{de Vaucouleurs1964}, and later supported by infrared photometry \citep[e.g.,][]{Blitz1991, Weiland1994, Dwek1995, Binney1997}, and asymmetries in star counts \citep[e.g.,][]{Nakada1991, Whitelock1992, Weinberg1992, Nikolaev1997, Stanek1995, Stanek1997}, as well as the post-Gaia 3D maps, which have clearly shown the bar in density maps \citep[e.g.,][]{Bailer-Jones2021, Bailer-Jones2023, Anders2019, Anders2022}.
However, its fundamental parameters, such as the mass, length, orientation with respect to the Sun, pattern speed, and the presence of a secondary bar, are currently difficult to measure directly, mainly because of the high dust extinction in the inner Galaxy and the highly complex of the kinematics and structure of the bar and/or bulge \citep[e.g.,][]{CabreraLavers2008, Nataf2010, McWilliam2010, Ness2016, Williams2016, Portail2017b}, owing to which researchers tend to use indirect methods to measure them.
One of the most effective methods is to analyze the properties of the outer Galactic disk stars that are strongly affected by resonant interactions of the bar \citep[e.g.,][]{Kalnajs1991, Dehnen1998, Dehnen2000, Antoja2014, Hunt2018}.

Kalnajs ({\color{blue}{1991}}) traced and measured the properties of the Galactic bar by the kinematic structures in the solar neighborhood, and confirmed the kinematics of the Sirius and Hyades moving groups may be related to the outer Lindblad resonance (OLR) of the Galactic bar.

Several studies after Kalnajs ({\color{blue}{1991}}) have used the local kinematic structures in the solar neighborhood to better trace and constrain the properties of the Galactic bar \citep[e.g.,][]{Dehnen2000, McMillan2013, Antoja2014}.
One of the most noteworthy kinematic structures is the Hercules streams \citep[e.g.,][]{Eggen1958, Dehnen1998, Antoja2014}, which is widely considered to be caused by the OLR of the Galactic bar \citep[e.g.,][]{Dehnen2000, Monari2017, Hunt2018}.
Based on the studies of these local kinematic structures, researchers constrain the measurement of the nature of the Galactic bar \citep[e.g.,][]{Dehnen2000, Antoja2014, Monari2017}.
However, these kinematic structures that may be related to the resonance of the Galactic bar are limited to the solar neighborhood, which inevitably leads to other possible origins.
Some authors have also suggested other origins of these local kinematic structures \citep[e.g.,][]{Antoja2009, Antoja2011}, for example, a moving group with a common birthplace within the disk, or the merger events, can also well explain the Hercules streams \citep[][]{Dehnen1998, Famaey2005, Bensby2007}.
Therefore, it is necessary to establish properties that are present over a large area of the disk to trace the resonance of the Galactic bar.
This is of great significance for the further constraint of the nature of the Galactic bar.

It is widely acknowledged that a rotating galactic bar has three fundamental resonances:
the corotation resonance, the inner Lindblad resonances, and the OLR \citep[e.g.,][]{Dehnen2000, Hunt2018}.
In either case, the features of the resonance of the bar should be present over a large area of the disk, and the effect of the resonance of the bar for the disk should be sustained, as well as both young/low-[$\alpha$/Fe] and old/high-[$\alpha$/Fe] populations should be affected by the resonance of the bar.
Therefore, we have to admit the fact that the resonance of the Galactic bar inevitably leads to periodic changes in the chemical and kinematic properties of the Galactic radius.
In this paper, we attempt to use the LAMOST-APOGEE red clump (RC) stars \citep[e.g.,][]{Bovy2014, Huang2020} to search for the periodic variation signals of chemical and kinematic properties with the Galactic radius, across a larger disk volume, thereby establishing the chemical and kinematic tracing for the signals of the resonance of the Galactic bar.

This paper is structured as follows.
In Section\,2, we briefly describe the data used in this study.
Our main results are presented and discussed in Section\,3.
Finally, we summarize our main conclusions in Section\,4.

\section{Data}

\begin{figure}[t]
\begin{center}
\includegraphics[width=8.9cm]{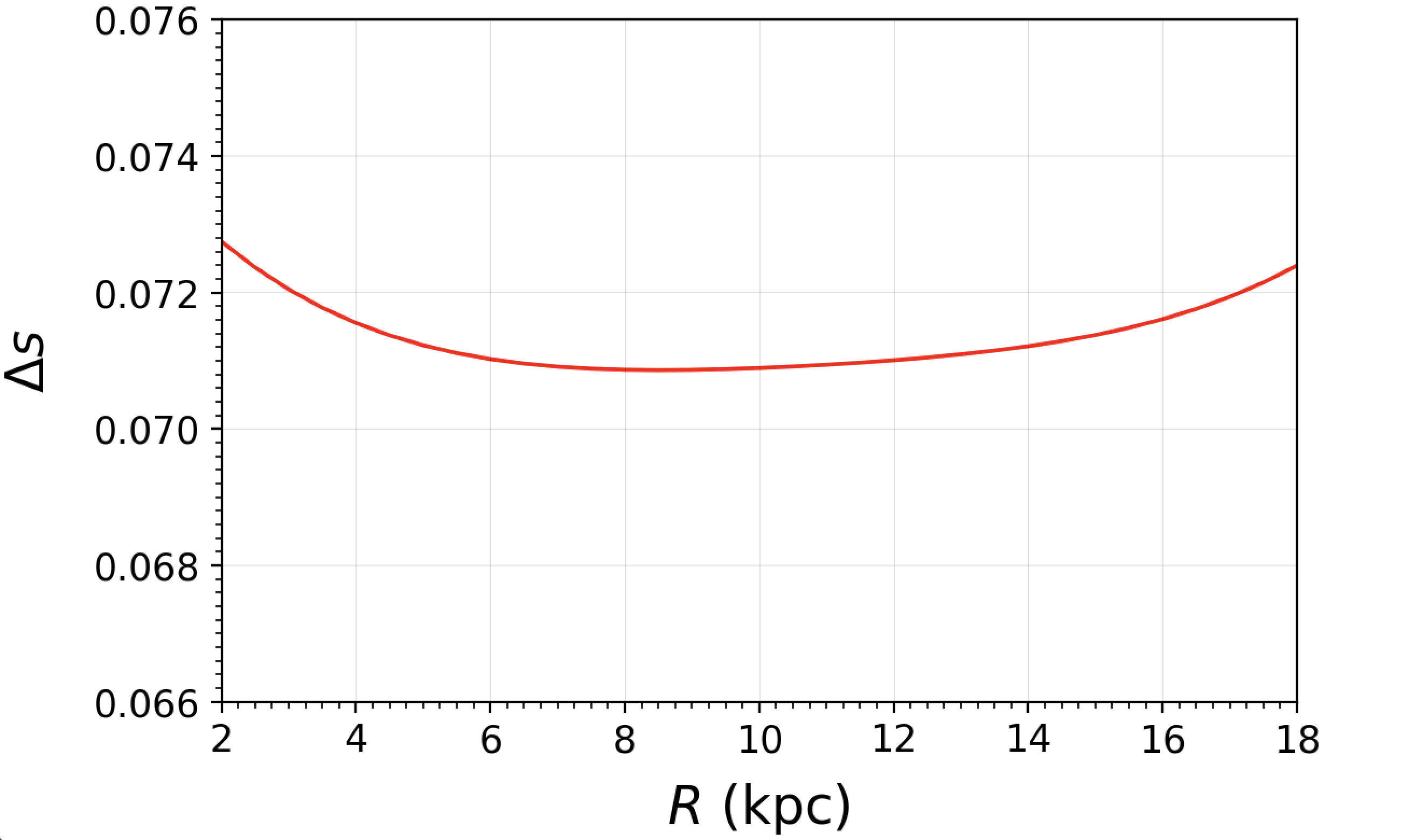}
\caption{The relative distance error ($\Delta s$, defined as the ratio of the distance error $\sigma d$ to distance $d$) as a function of $R$.
The red line is the best fit for the whole RC sample stars.}
\end{center}
\end{figure}

Standard Galactocentric cylindrical coordinates ($R$,\,$\phi$,\,$Z$) are used in this study, with three velocity components, $V_{R}$, $V_{\phi}$ and $V_{Z}$, respectively.
To estimate the positions and kinematics, we set the Galactocentric distance of the Sun as $R_{0}$ = 8.34\,kpc \citep{Reid2014}, local circular velocity as $V_{c, R_{0}}$ = 238.0\,km\,s$^{-1}$ \citep[e.g.,][]{Reid2004, Schonrich2010, Schonrich2012, Reid2014, Huang2015, Huang2016, Bland-Hawthorn2016}, and the solar motions as ($U_{\odot}$, $V_{\odot}$, $W_{\odot}$) $=$ $(13.00, 12.24, 7.24)$\,km\,s$^{-1}$ \citep{Schonrich2018}.

In this study, we mainly use a sample with 177,123 RC stars \citep[e.g.,][]{Bovy2014, Huang2020} from the LAMOST \citep[e.g.,][]{Deng2012, Cui2012, Liu2014, Yuan2015} and APOGEE \citep{Majewski2017} surveys, where 137,448 and 39,675 RC stars, respectively, are from LAMOST \citep{Huang2020} and APOGEE \citep{Bovy2014} surveys.
For the common sources, we selected the parameters with a large signal-to-noise ratio (SNR) of the two datasets.
The uncertainties of the surface gravity (log\,$g$), effective temperature ($T_{\rm eff}$), line-of-sight velocity ($V_{\rm r}$), metallicity ([Fe/H]) and [$\alpha$/Fe] for the whole LAMOST-APOGEE RC sample are typically better than 0.10\,dex, 100\,K, 5\,km\,s$^{-1}$, 0.10$-$0.15\,dex and 0.03$-$0.05\,dex, respectively \citep[see][]{Bovy2014, Huang2020}.
The distance of each star is determined by its standard candle nature, with uncertainty typically around 5\%-10\%, and the relative distance error ($\Delta s$, defined as the ratio of the distance error $\sigma d$ to distance $d$) as a function of $R$ is shown in Fig.\,1.
To improve the accuracy of the kinematic and dynamic calculations, we use Gaia DR3 \citep[e.g.,][]{Gaia Collaboration2023a, Gaia Collaboration2023b, Recio-Blanco2023} to update the astrometric parameters (e.g., proper motions) of the whole sample stars.

\begin{figure}[t]
\begin{center}
\includegraphics[width=8.6cm]{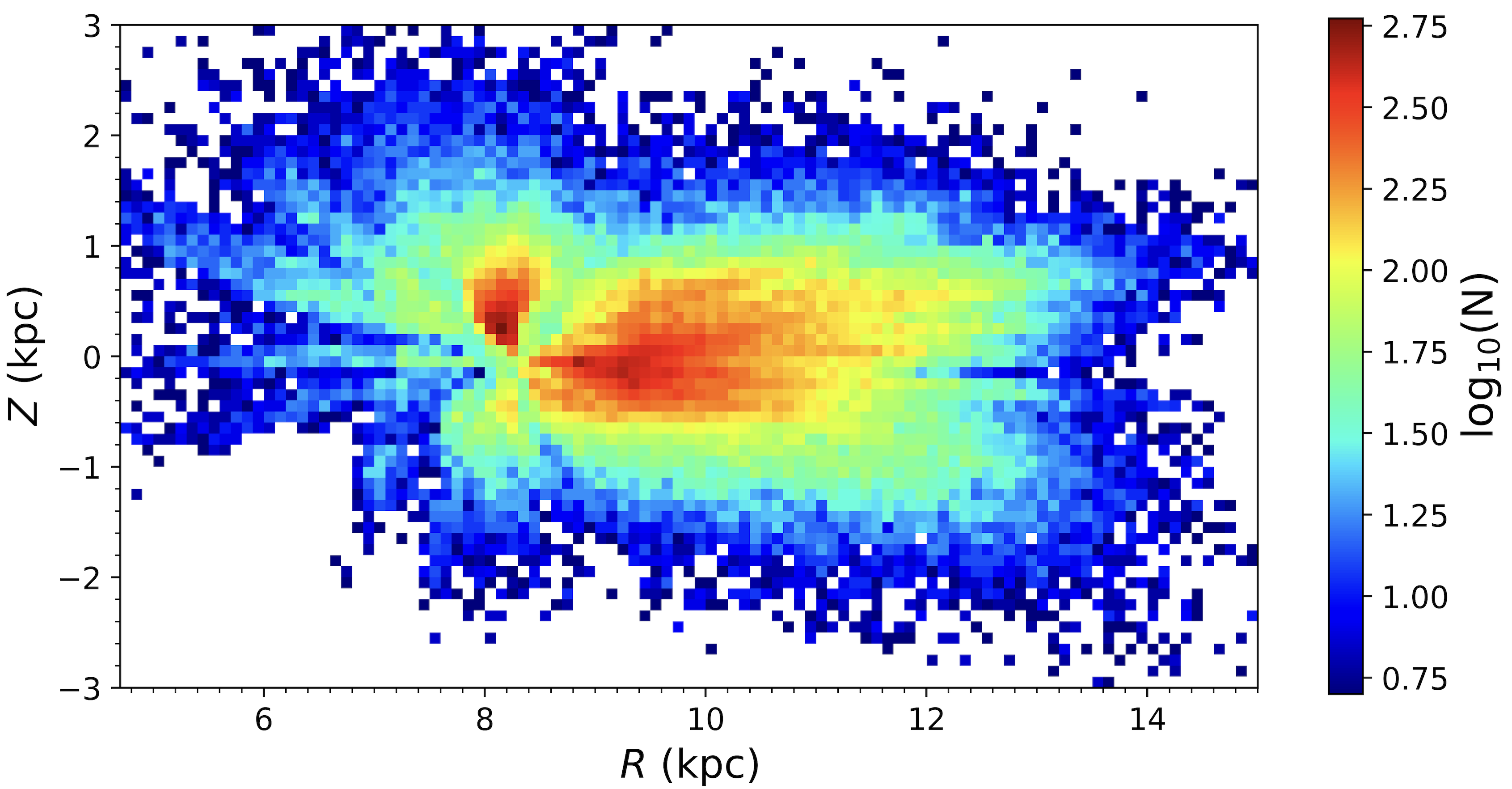}
\caption{Spatial distribution of the LAMOST-APOGEE RC stars in the $R$ - $Z$ plane, with color-coded by the stellar number densities.
Both axes are spaced by 0.1\,kpc, with no less than 8 stars in a bin.}
\end{center}
\end{figure}

\begin{figure}[t]
\begin{center}
\includegraphics[width=8.4cm]{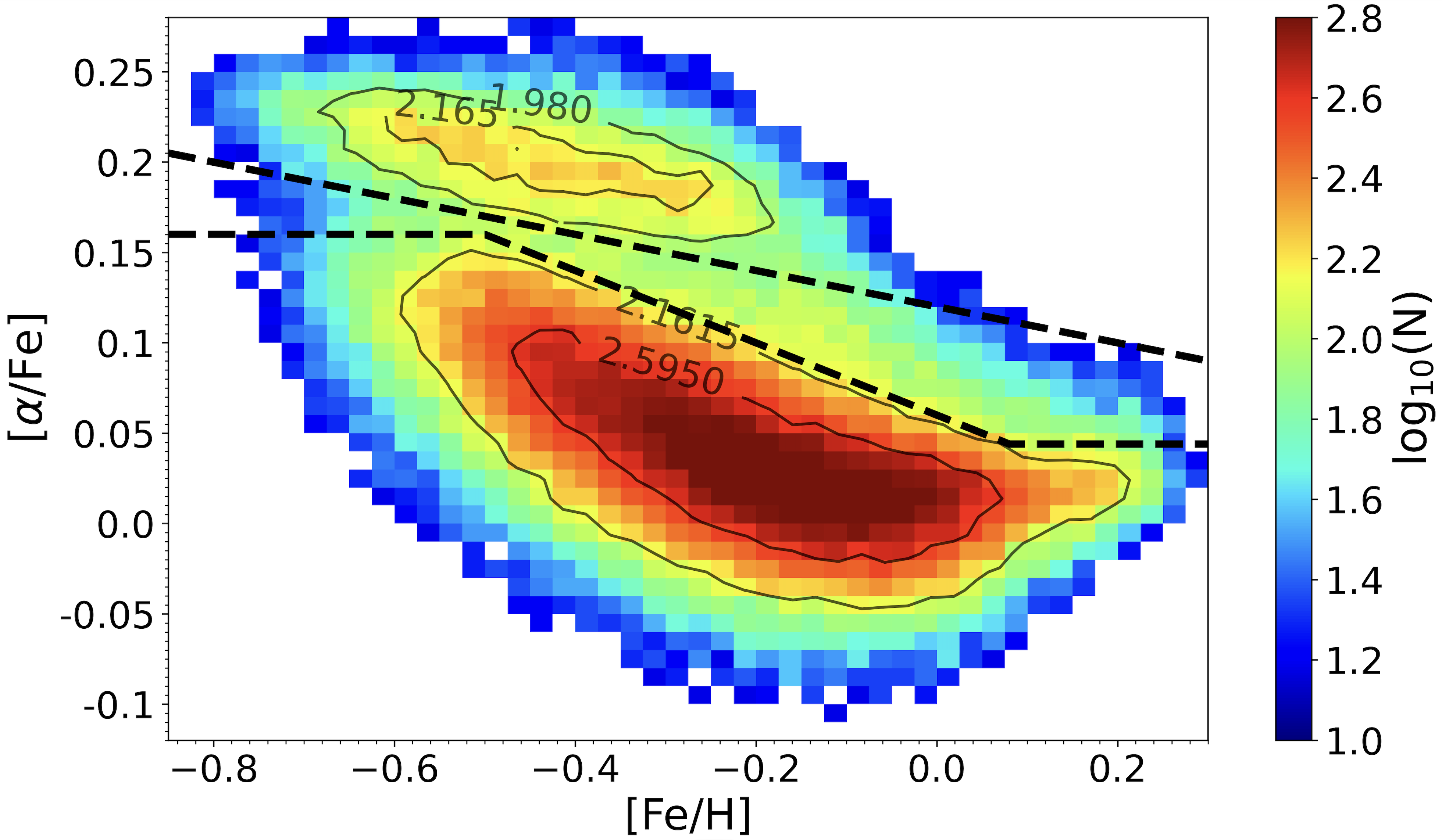}
\caption{The [Fe/H]$-$[$\alpha$/Fe] distribution of the LAMOST-APOGEE RC stars, with color-coded by the logarithmic number densities, and overplotted with contours of equal densities.
The [Fe/H] and [$\alpha$/Fe] axes are spaced by 0.025\,dex and 0.02\,dex, respectively.
There is a minimum of 20 stars per bin.
The two dashed lines are used to separate the thin (below the lines) and thick (above the lines) disks.}
\end{center}
\end{figure}

The velocity dispersions ($\sigma$) in each spatial bin are estimated by 3$\sigma$ clipping to remove outliers.
To ensure a reliable Galactic disk analysis, the conditions with [Fe/H] $\geq -1.0$\,dex and $|V_{Z}|$ $\leq$ 120\,km\,s$^{-1}$ are further set for excluding the halo stars \citep[e.g.,][]{Huang2018, Hayden2020, Sun2020, Sun2023}.
Finally, 170,729 RC stars are eventually selected, of which 39,112 and 131,617 RC stars respectively from APOGEE and LAMOST surveys.
The spatial distribution of the final selected stars is shown in Fig.\,2.

Since the [Fe/H] and [$\alpha$/Fe] are typically distinguished for the two datasets, we further calibrate these parameters of the APOGEE dataset to the LAMOST dataset based on the best corrections obtained from the random forest model of the machine-learning method trained with the common stars of the two datasets.
The [Fe/H]--[$\alpha$/Fe] relation of the whole sample stars with calibrated datasets is shown in Fig.\,3.
The plot shows a bimodal distribution of two branches, with the high-[$\alpha$/Fe] and low-[$\alpha$/Fe] populations corresponding to thick and thin disks, respectively.
To get pure thin and thick disk stars, two empirical cuts (see Fig.\,3) are used to separate the two populations based on previous work \citep[e.g.,][]{Bensby2005, Lee2011, Brook2012, Haywood2013, Recio-Blanco2014, Nidever2014, Guiglion2015, Hayden2015, Queiroz2020, Sun2024a}.
With the cuts, we selected 135,009 thin disk stars (below the lower cut) and 23,168 thick disk stars (above the upper cut), and the fundamental properties of the two disks are detailedly summarized in Table 1.

\begin{table*}
\caption{The properties of the thin/thick disks}

\centering
\setlength{\tabcolsep}{6mm}{
\begin{tabular}{lllllllll}
\hline
\hline
\specialrule{0em}{5pt}{0pt}
Name                                         & $\langle$ [$\alpha$/Fe] $\rangle$  &  $\langle$ [Fe/H] $\rangle$  &     $\left\langle R \right\rangle$   &          $\left\langle V_{\phi} \right\rangle$        &  $\sigma_{\phi}$  &     Number     &    Fraction \\
                                             &                (dex)               &        (dex)       &     (kpc)     &        (km\,s$^{-1}$)           &    (km\,s$^{-1}$)    &           &               \\
\specialrule{0em}{5pt}{0pt}
\hline
\specialrule{0em}{3pt}{0pt}
Thin disk   &  0.03   &            $-$0.2             &     10.14     &            225.01               &          22.97       &    135,009 &      79.08\%\\ [0.2cm]
Thick disk  &  0.20   &            $-$0.4            &     8.54      &            179.27               &          54.81       &   23,168  &      13.57\%   \\
\specialrule{0em}{3pt}{0pt}
\hline
\specialrule{0em}{3pt}{0pt}
\end{tabular}}
\label{tab:datasets}
\end{table*}

\begin{figure*}[t]
\centering
\subfigure{
\includegraphics[width=8.4cm]{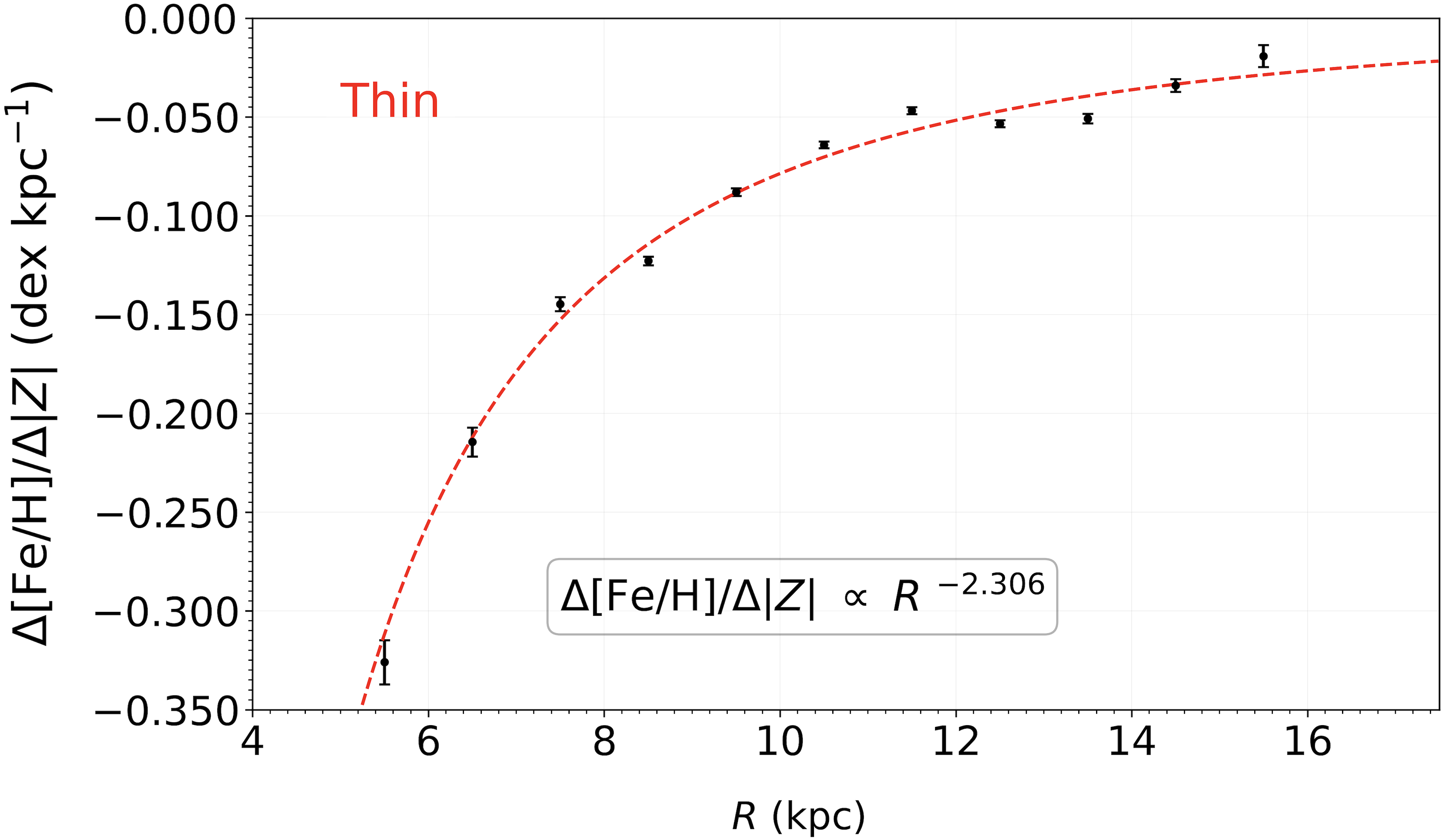}
}
\hspace{0.5cm}
\subfigure{
\includegraphics[width=8.4cm]{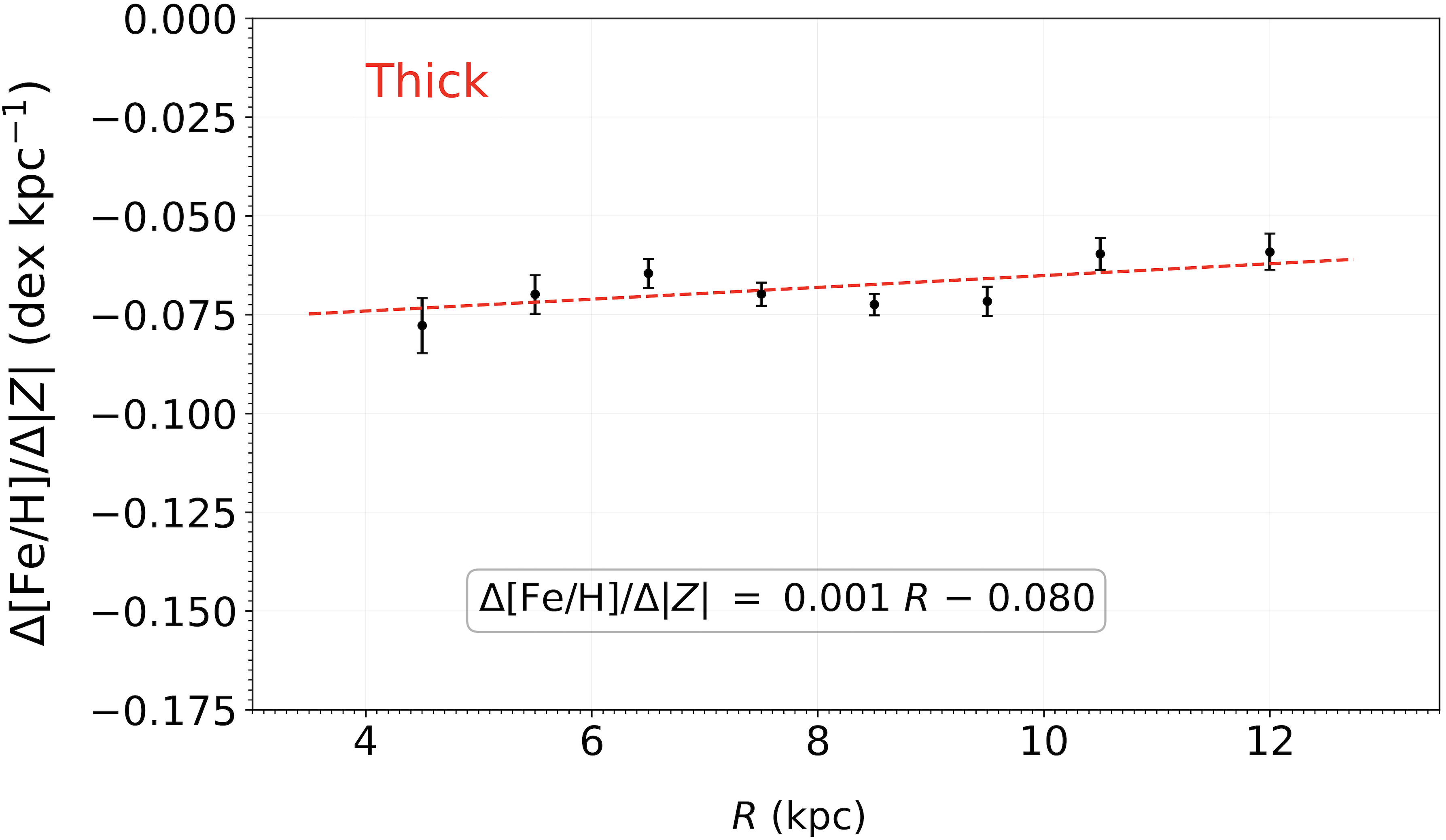}
}

\caption{The $\Delta$[Fe/H]/$\Delta|Z|$ as a function of the $R$ for the thin (left panel) and thick (right panel) disks.
The red dashed lines are the best fit of the global profiles of trends of $\Delta$[Fe/H]/$\Delta|Z|$--$R$.}
\end{figure*}

\section{Results and discussion}
The fact that the resonance from the rotating Galactic bar should be sustained, and its impact on the disk should be present over a larger area, as well as both young/thin and old/thick disks should be affected by it.
Therefore, the resonance of the Galactic bar should induce detectable periodic changes with $R$ for the chemical and kinematic properties.

As mentioned above, the RC stars are used for this study, which are widely distributed across the entire Galactic disk, and hence are excellent tracers to study the chemical, kinematic and dynamic properties of the Galactic disk \citep[e.g.,][]{Huang2016, Huang2020, Sun2020, Sun2024a}.
Leveraging this sample, any chemical and kinematic signals of the resonance of the Galactic bar should be clearly detected.

\subsection{The chemical signature of the resonance of the Galactic bar}

\begin{figure*}[t]
\centering
\subfigure{
\includegraphics[width=17.9cm]{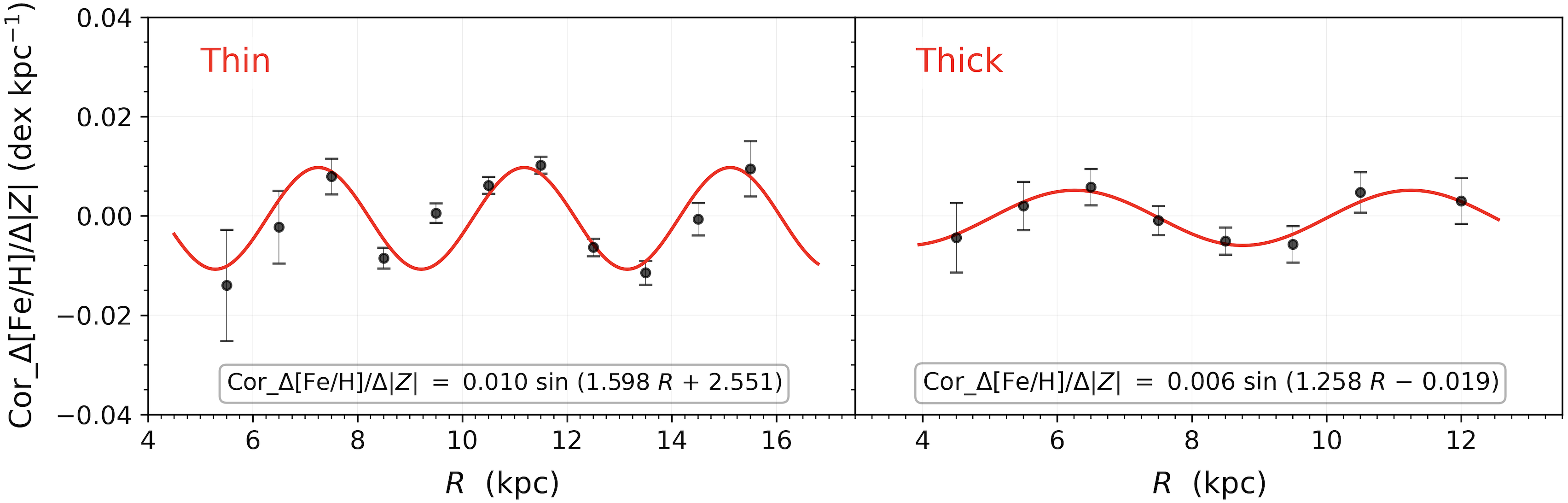}
}

\caption{The vertical metallicity gradient with subtracted the fitted profile in Fig.\,4, Cor\,$\Delta$[Fe/H]/$\Delta|Z|$ (defined as $\Delta$[Fe/H]/$\Delta|Z|$ $-$ Fit\_$\Delta$[Fe/H]/$\Delta|Z|$), as a function of $R$, for thin (left panel) and thick (right panel) disks.
The red lines are the best fit with Equation (1).}
\end{figure*}

To detect the periodic changes with $R$ for the chemical properties, we examine the vertical metallicity gradients ($\Delta$[Fe/H]/$\Delta$Z) as a function of $R$, of the thin and thick disks, and the results are shown in Fig.\,4.

The $\Delta$[Fe/H]/$\Delta|Z|$ of the thin disk displays an increasing trend as $R$ increases and, showing a single power law shape, increasing steadily from $-$0.36\,dex\,kpc\,$^{-1}$ at $R$ $\sim$ 5.5\,kpc to $-$0.05\,dex\,kpc\,$^{-1}$ at $R$ larger than 11.5\,kpc (see left panel of Fig.\,4).
The thick disk shows an almost constant value (nearly $-$0.06 $\sim$ $-$0.08\,dex\,kpc\,$^{-1}$) with radius (see right panel of Fig.\,4).
Our results in the solar neighborhood (8.0 $\leq$ $R$ $<$ 9.0\,kpc) indicate that the thin and thick disks respectively display $-$0.1228 $\pm$ 0.0021 dex\,kpc$^{-1}$ and $-$0.0724 $\pm$ 0.0027 dex\,kpc$^{-1}$ (see Fig.\,4).
This result is in good agreement with the LAMOST A/F/G/K-type giant sample of Yan et al. ({\color{blue}{2019}}), who reported the thin and thick disks stars with 8.0 $\leq$ $R$ $<$ 9.0 kpc are $\Delta$[Fe/H]/$\Delta|Z|$ = $-$0.1237 $\pm$ 0.0017 dex\,kpc$^{-1}$ and $\Delta$[Fe/H]/$\Delta|Z|$ = $-$0.0705 $\pm$ 0.0017 dex\,kpc$^{-1}$, respectively.
Our results in the outer solar neighborhood (see Fig.\,4) are obviously different from Yan et al. ({\color{blue}{2019}}), while it is in rough agreement with the LAMOST RC samples with $|Z|$ $\leq$ 1.0\,kpc of Huang et al. ({\color{blue}{2015}}).
Since the distance of the sample of Yan et al. ({\color{blue}{2019}}) is determined from Gaia Parallax, their results may lead to larger uncertainties compared to ours.

The global trends of the $\Delta$[Fe/H]/$\Delta|Z|$ with $R$ in our results are significant (see Fig.\,4).
As mentioned above, the thin and thick disks form shapes of a global single power law and a global liner function, respectively.
These global trends could be attributable to a broad variety of factors, such as the stellar migrations of the thin and thick disks \citep[e.g.,][]{Sellwood2002, Schonrich2009, Loebman2011, Schonrich2017, Yan2019, Sun2024b}; flaring of the thin disk \citep[e.g.,][]{Hunter1969, Khoperskov2017, Sun2020, Vickers2021}, and warping of the thin (perhaps also of the thick) disk \citep[e.g.,][]{Momany2006, Poggio2018, Mackereth2019, Li2020, Sun2024a}.
However, we are not aiming at the origins of these global trends, but at tracing the signal of the resonance of the Galactic bar, and therefore, we attempt to subtract these global trends to make the signal of the resonance of the Galactic bar as easy to detect as possible.

The global profiles of the $\Delta$[Fe/H]/$\Delta|Z|$--$R$ of the thin and thick disk are fitted by the red dashed line in Fig.\,4.
The results yield the global profiles of the thin and thick disks are respectively $\Delta$[Fe/H]/$\Delta|Z|$ $\propto$ $R^{-2.306}$ and $\Delta$[Fe/H]/$\Delta|Z|$ = 0.001\,$R$\,$-$\,0.080 (to distinguish it from the original data, hereafter, we name these fitted $\Delta$[Fe/H]/$\Delta|Z|$ as Fit\_$\Delta$[Fe/H]/$\Delta|Z|$).
We can find that the original data points exhibit some oscillations with respect to the best-fit curves as increasing $R$ (see Fig.\,4).
After subtracting the fitted profile of the original data, we present the corrected vertical metallicity gradient (Cor\_$\Delta$[Fe/H]/$\Delta|Z|$ = $\Delta$[Fe/H]/$\Delta|Z|$ $-$ Fit\_$\Delta$[Fe/H]/$\Delta|Z|$) as a function of $R$ in Fig.\,5.
The results indicate that the Cor\_$\Delta$[Fe/H]/$\Delta|Z|$ displays some significant oscillations with $R$ for the thin and thick disk, with the profiles displaying obvious sine-function shapes.
Therefore, we fit these profiles with a sine-function as follows:

\begin{equation}
\label{eq:tiltangle2}
    \mathrm{Cor\_\Delta[Fe/H]/\Delta|Z|} = A\,\mathrm{sin}(\omega R + \Phi)
\end{equation}
The best fit is displayed by the red line in each figure, the results of those yield Cor\_$\Delta$[Fe/H]/$\Delta|Z|$ = 0.010\,$\mathrm{sin}$\,(1.598\,$R$\,+\,2.551) and Cor\_$\Delta$[Fe/H]/$\Delta|Z|$ = 0.006\,$\mathrm{sin}$\,(1.258\,$R$\,$-$\,0.019) for thin and thick disks, respectively.
Both thin and thick disks exhibit periodic oscillations in Cor\_$\Delta$[Fe/H]/$\Delta|Z|$ with $R$, which indicates that the resonances of the Galactic bar are the most possible reason for these oscillations.
In addition, the phases of the fitted sine-functions are distinguished for the thin and thick disks (see Fig.\,5), which indicates the two populations have different resonance modes, and therefore, we consider this result likely to imply the presence of a secondary bar in the Milky Way \citep[e.g.,][]{Portail2017a}.

\subsection{The kinematic signature of the resonance of the Galactic bar}

The resonance of the Galactic bar would cause perturbations in the stellar orbits at locations of the principal resonances \citep[e.g.,][]{Collett1997}, thereby forming observable structures/velocity groups \citep[e.g.,][]{Dehnen2000, Antoja2014, Hunt2018, Trick2021}.
However, the Galactic disk might have undergone long-term dynamical heating \citep[e.g.,][]{Spitzer1953, Barbanis1967, Sun2023} and/or multiple disturbance events \citep[e.g.,][]{Quinn1993, Abadi2003, Brook2004, Sun2023}.
This should lead to the structures/velocity groups to undergo dynamic evolution with time, thereby making it difficult for us to capture the complete resonance signal in a larger space.
It is also difficult to rule out whether the velocity groups are formed from some disturbance events of the Galactic disk, such as merger and accretion \citep[e.g.,][]{Dehnen1998, Bensby2007}.
Therefore, we consider using the analysis of a kinematic feature that can be powerful to detect the gravitational potential of the Galactic disk, and it is sensitive to the disturbance of gravitational potential by the Galactic bar.
In this view, we attempt to analyze the tilt of the velocity ellipsoid features.

The orientation of the velocity ellipse is calculated following Smith et al. ({\color{blue}{2009}}), that is, for observed two orthogonal velocity components, $v_{i}$ and $v_{j}$, of a group stars, the tilt angle $\alpha_{ij}$ is defined as:

\begin{equation}
\label{eq:tiltangle1}
    \tan(2 \alpha_{ij}) = \frac{2 \mathrm{cov}(v_i, v_j)}{\mathrm{var}(v_i) - \mathrm{var}(v_j)} = \frac{2\sigma_{ij}^{2}}{\sigma_{ii}^{2} - \sigma_{jj}^{2}}
\end{equation}
where

\begin{equation}
\mathrm{cov}(v_i, v_j) \equiv \sigma^2_{ij} \equiv
\left\langle
(v_i - \langle v_i \rangle)(v_j - \langle v_j \rangle)
\right\rangle
\label{eq:cov}
\end{equation}
and the angled brackets are the averaging over the phase-space distribution function \citep[see e.g.,][]{Binney2008, Smith2009, Sun2023}.

\begin{figure*}[t]
\centering
\subfigure{
\includegraphics[width=17.5cm]{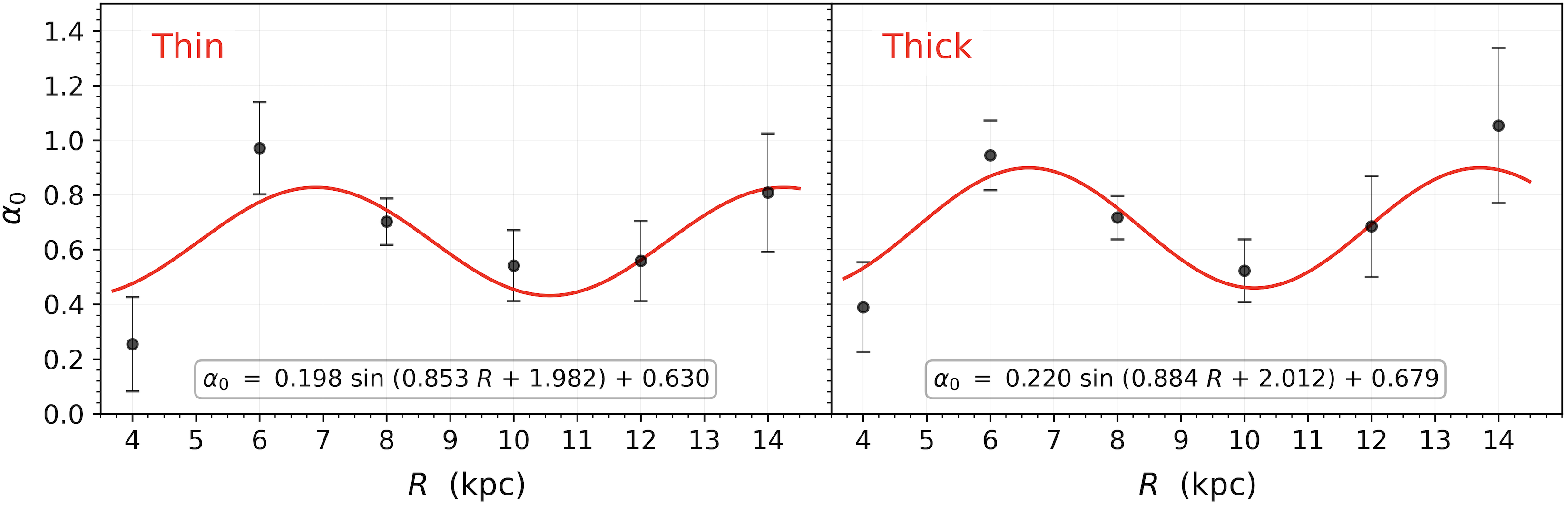}
}

\caption{The $\alpha_{0}$ as a function of $R$ for thin (left panel) and thick (right panel) disks.
The red lines are the best fit with Equation (5).}
\end{figure*}

We measure the tilt of the velocity ellipsoid in the $R$--$Z$ phase space for the whole RC sample stars at $|Z|$ $\leq$ 3.0\,kpc and 4.0 $\leq$ $R$ $\leq$ 15.0\,kpc, with binned by 0.5\,kpc in both axes, and only estimate the bins with no less than 10 stars.
Then, we summarize the tilt angles of the velocity ellipsoids for various $R$ regions, with a simple model following previous studies \citep[e.g.,][]{Binney2014, Budenbender2015}, that is,

\begin{equation}
\label{eq:tiltangle2}
    \alpha = \alpha_{0}\, \mathrm{arctan}(\frac{Z}{R})
\end{equation}
where $\alpha_{0}$ is the fitting constant.
A result of $\alpha_{0}$ = 1.0 indicates exact spherical alignment, whilst $\alpha_{0}$ $<$ 1.0 indicates the ellipsoids tend to tilt towards cylindrical alignment.
The tilt of the velocity ellipsoid towards cylindrical alignment may be related to the gravitational potential of the baryonic disk \citep[e.g.,][]{Hagen2019, Everall2019, Sun2023}.
Therefore, the $\alpha_{0}$ should be present periodic changes with $R$ if the gravitational potential of the Galactic disk is affected by the resonance of the Galactic bar.

The $\alpha_{0}$ as a function of $R$ for the thin and thick disks are shown in Fig.\,6.
The results indicate that the $\alpha_{0}$ displays oscillations with $R$ for both thin and thick disks.
At $R$ = 8.0\,kpc, our results display $\alpha_{0}$ $\sim$ 0.703 $\pm$ 0.085 and $\alpha_{0}$ $\sim$ 0.717 $\pm$ 0.079 for thin and thick disks, respectively.
This is in agreement with the results of Sun et al. ({\color{blue}{2023}}).
The profiles of the $\alpha_{0}$--$R$ of the thin and thick disks also display obvious sine-function shape, and therefore, we fit the profiles with a function similar to equation (1), that is,

\begin{equation}
\label{eq:tiltangle2}
    \alpha_{0} = A\,\mathrm{sin}(\omega R + \Phi) + C
\end{equation}
The best fits are displayed by the red line in the figure, the results yield $\alpha_{0}$ = 0.198\,$\mathrm{sin}$\,(0.853\,$R$\,+\,1.982)\,+\,0.630 and $\alpha_{0}$ = 0.220\,$\mathrm{sin}$\,(0.884\,$R$\,+\,2.012)\,+\,0.679 for thin and thick disks, respectively.
This is in rough agreement with the result of LAMOST data in Sun et al. ({\color{blue}{2023}}).
The periodic oscillations in $\alpha_{0}$ with $R$ mean the perturbations of the gravitational potential of the baryonic disk display periodic oscillations with $R$.
Since these oscillations appear in both thin and thick disks, we can further confirm that the resonance of the Galactic bar is likely one of the most dynamic origins.
In addition, from the phases of the fitted sine-functions, we can further confirm our result in Sec.3.1, that is, there is a secondary bar in the Milky Way \citep[e.g.,][]{Portail2017a}

\section{Conclusions}

In this paper, we use over 170,000 RC stars selected from LAMOST and APOGEE surveys to detect the chemical and kinematic signatures of the resonances of the Galactic bar, over a wide range of Galactocentric radii.
We find that:
\\
\\
$\bullet$ The relation of $\Delta$[Fe/H]/$\Delta|Z|$--$R$ with subtracted the profiles trends, displays clear oscillations in Cor\_$\Delta$[Fe/H]/$\Delta|Z|$ with $R$ for both thin and thick disks.
The thin and thick disks values are Cor\_$\Delta$[Fe/H]/$\Delta|Z|$ = 0.010\,$\mathrm{sin}$\,(1.598\,$R$\,+\,2.551) and Cor\_$\Delta$[Fe/H]/$\Delta|Z|$ = 0.006\,$\mathrm{sin}$\,(1.258\,$R$\,$-$\,0.019), respectively.
\\
\\
$\bullet$ The tilt angle of the velocity ellipsoid is accurately described as $\alpha$ = $\alpha_{0}$\,\,arctan\,(Z/R), and $\alpha_{0}$ also displays oscillations with $R$ for both thin and thick disks.
The results indicate that $\alpha_{0}$ = 0.198\,$\mathrm{sin}$\,(0.853\,$R$\,+\,1.982)\,+\,0.630 and $\alpha_{0}$ = 0.220\,$\mathrm{sin}$\,(0.884\,$R$\,+\,2.012)\,+\,0.679 for thin and thick disks, respectively.
\\
\\
These periodic oscillations in Cor\_$\Delta$[Fe/H]/$\Delta|Z|$ and $\alpha_{0}$ with $R$ appear in both thin and thick disks, are the most likely chemical and kinematic signatures of the resonance of the Galactic bar.
The difference in the phase of the fitted sine-functions of the periodic oscillations for the thin and thick disks may be related to the presence of a second Galactic bar.

\section*{Acknowledgements}
We thank the anonymous referee for very useful suggestions to improve the work.
This work is supported by the NSFC projects 12133002, 12073070, and 12003027, and National Key R\&D Program of China No. 2019YFA0405503, and CMS-CSST-2021-A09.

Guoshoujing Telescope (the Large Sky Area Multi-Object Fiber Spectroscopic Telescope LAMOST) is a National Major Scientific Project built by the Chinese Academy of Sciences. Funding for the project has been provided by the National Development and Reform Commission. LAMOST is operated and managed by the National Astronomical Observatories, Chinese Academy of Sciences. The LAMOST FELLOWSHIP is supported by Special Funding for Advanced Users, budgeted and administrated by Center for Astronomical Mega-Science, Chinese Academy of Sciences (CAMS)

\bibliographystyle{aasjournal}

\end{document}